\documentclass[prb,letterpaper,twocolumn,showpacs,preprintnumbers,citeautoscript,floatfix,amsmath,amssymb]{revtex4-1}

\usepackage{epsfig}
\usepackage{graphicx}
\usepackage{color}
\usepackage{soul}
\usepackage{amsfonts}

\begin{document}

\title{Structural transitions and transport-half-metallic ferromagnetism in LaMnO$_3$ at elevated pressure}

\author{Jiangang He$^{1}$}
\author{ Ming-Xing Chen$^{2}$}
\author{Xing-Qiu Chen$^{3}$}
\author{Cesare Franchini$^{1,3}$}
\email[Corresponding author: ]{cesare.franchini@univie.ac.at}
\affiliation{$^{1}$Faculty of Physics, University of Vienna and Center for Computational
 Materials Science, A-1090 Wien, Austria}
\affiliation{$^{2}$Faculty of Chemistry, University of Vienna and Center for Computational
Materials Science, A-1090 Wien, Austria}
\affiliation{$^{3}$Shenyang National Laboratory for Materials Science, Institute of Metal Research, 
Chinese Academy of Sciences, Shenyang 110016, China}

\date{\today}


\begin{abstract}
By means of hybrid density functional theory we investigate the evolution of the structural, electronic 
and magnetic properties of the colossal magnetoresistance (CMR) parent compound LaMnO$_3$ under pressure.
We predict a transition from a low pressure antiferromagnetic (AFM) insulator to a 
high pressure ferromagnetic (FM) transport half-metal (tHM), characterized by a large spin polarization 
($\approx$ 80-90 \%).
The FM-tHM transition is associated with a progressive quenching of the cooperative
Jahn-Teller (JT) distortions which transform the $Pnma$ orthorhombic phase into a perfect cubic one
(through a mixed phase in which JT-distorted and regular MnO$_6$ octahedra coexist),
and with a high-spin ($S=2$, $m_{Mn}$=3.7 $\mu_{\rm B}$) to low-spin ($S=1$, $m_{Mn}$=1.7 $\mu_{\rm B}$) 
magnetic moment collapse. These results interpret the progression of the experimentally observed 
non-Mott metalization process and open up the possibility of realizing CMR behaviors in a stoichiometric 
manganite.
\end{abstract}

\maketitle

\section{Introduction}

Half-metallic ferromagnets (HMFs) are magnetic compounds that are simultaneously metallic
and insulator, depending on the spin channel\cite{Groot83,Katsnelson08}. Their ability to 
provide fully spin-polarized currents make them ideal candidates for spintronic
applications\cite{Zutic04,Prinz98}. A prominent class of HMFs is represented by 
strongly correlated CMR manganites ($\rm La_{1-x}$$A_x$$\rm MnO_3$,
$A$=Ca, Sr, or Ba)\cite{Coey97}: hole-doping $\rm LaMnO_3$ through the 
substitution of La with $A$=Ca, Sr, or Ba creates itinerant holes that
progressively lead to an antiferromagnetic (AFM)-insulator to ferromagnetic (FM)-metal 
transition and critically determine the coexistence of half-metallic spin imbalance and the 
so called CMR effect, i.e. a dramatic change of the electrical resistance in the presence 
of a magnetic field\cite{Jonker50, Salamon01}.
Soon after the pioneering observation of CMR effect by Jonker and van Santen\cite{Jonker50},
Zener\cite{Zener51} proposed the double exchange (DEX) mechanism to explain the          
AFM-insulator ($x=0$) to FM-metal ($0.2<x<$0.5) transition, relaying on the O$^{2-}$-mediated transfer 
of an electron between inequivalent Mn$^{3+}$ and Mn$^{4+}$ sites.
The DEX mechanism qualitatively explained the ferromagnetic interactions and the observed metallic 
behavior below the Curie temperature T$_{\rm C}$, but it turned out to be inadequate to explain the
observed high insulating-like resistivity above the transition temperature and, even most importantly, 
the observation of CMR in stoichiometric phases like Tl$_2$Mn$_2$O$_7$\cite{Shimakawa96}.
Conversely, the more recent half-metallic ferromagnetic model\cite{Pickett96, Nadgorny01}, 
which is based on the spin-polarized calculation of the density of states (DOS) within the Density 
Functional Theory (DFT), provides several clues to the underlying processes involved in the CMR 
phenomena and since it does not rely on the mixed-valence (Mn$^{3+}$ and Mn$^{4+}$) picture it can 
explain the observation of CMR in stoichiometric phases.\cite{Shimakawa96} 

Controlling and understanding CMR-HMFs phenomena in manganites within the DFT framework remains 
a great challenge because of two fundamental obstacles: 
(i) strong exchange-correlation effects and the concurrent orbital/lattice/spin correlations which are 
not well described by conventional DFT methods, and (ii) doping-induced structural disorder, which 
unavoidably limits the application of quantum mechanical schemes based on repeated unit cells. 
The drawbacks of DFT in dealing with insulating transition metal oxides can be corrected by employing more
sophisticated methods such us hybrid functionals\cite{Becke93} which have been proven to provide 
substantially improved structural, electronic and magnetic properties\cite{Munoz04, Archer2011, Franchini2010, 
Franchini2011}, thanks to the inclusion of a portion of exact 'non-local' exchange. Beyond-DFT approaches are particularly 
necessary to correctly predict structural distortions, magnetic energies and bandgap in LaMnO$_3$, which are 
wrongly described by DFT\cite{Munoz04,Trimarchi05,Hashimoto10}: DFT is unable to describe the 
JT instabilities, and stabilize a metallic FM solution instead of the experimentally observed JT-distorted 
AFM insulating state\cite{Hashimoto10}.
As for the structural disorder, the possibility to circumvent the problem by realizing CMR-HMF behaviors
in stoichiometric samples would represent a substantial benefit not only for theory, but also for the 
experimental and technological research, thanks to the higher degree of control and manipulability of the 
relevant physical processes. However the task is hard, and up-to-date very few stoichiometric
CMR compounds have been identified\cite{Shimakawa96, Ishiwata2011}.

Boosted by the recent experimental observations of an insulator-to-metal transition (IMT)
in dense LaMnO$_3$\cite{Loa01, Baldini11} at about 32 GPa, in this article we explore 
the structural and magnetoelectric response of LaMnO$_3$ upon compression up to 150 GPa. 

\begin{table*}[t]
\caption{Structural parameters of LaMnO$_3$: low-temperature (4.2 K)
experimental data\cite{elemans71} versus fully optimized PBE and HSE
results with different values of the exact-exchange mixing parameter
$\alpha$: 0 (PBE),  0.10 (HSE-10), 0.15 (HSE-15), 0.25 (HSE-25), and
0.35 (HSE-35). Mn-O$_{s}$, Mn-O$_{m}$ and Mn-O$_{l}$, represent the
short, medium and long Mn-O bondlengths, whereas Mn-O$_m$-Mn
($^\circ$) and Mn-O$_{s/l}$-Mn ($^\circ$) indicates the
corresponding angles. Finally, the JT parameters Q$_2$ and Q$_3$ are
defined as: Q$_2=2(l-s)/\sqrt(2)$ and Q$_3=2(2m-l-s)/\sqrt(6)$. }
\vspace{0.3cm}
\begin{ruledtabular}
\begin{tabular}{cccccccc}
                      & Exp$^a$& HSE-35 & HSE-25 & HSE-15 & HSE-10 &  PBE    \\
Volume (\AA$^3$)      & 243.57 & 243.98 & 245.82 & 247.36 & 244.24 & 244.21  \\
\emph{a} (\AA)        & 5.532  &  5.526 & 5.537  &  5.553 &  5.661 &  5.569  \\
\emph{b} (\AA)        & 5.742  &  5.789 & 5.817  &  5.820 &  5.594 &  5.627  \\
\emph{c} (\AA)        & 7.668  &  7.628 & 7.633  &  7.653 &  7.712 &  7.793  \\
Mn-O$_m$ (\AA)        & 1.957  &  1.954 & 1.957  &  1.962 &  1.979 &  1.992  \\
Mn-O$_{l}$ (\AA)     & 2.184  &  2.204 & 2.214  &  2.213 &  2.134 &  2.063  \\
Mn-O$_{s}$ (\AA)     & 1.903  &  1.899 & 1.905  &  1.914 &  1.923 &  1.971  \\
Mn-O$_m$-Mn ($^\circ$)& 154.3  & 154.78 & 154.35 & 154.36 & 153.96 & 155.85  \\
Mn-O$_{s/l}$-Mn ($^\circ$)& 156.7  & 154.38 & 154.08 & 154.17 & 157.59 & 157.71  \\
Q$_2$                 & 0.398  &  0.431 & 0.437  &  0.423 &  0.298 &  0.131  \\
Q$_3$                 &-0.142  & -0.159 &-0.167  & -0.165 & -0.080 & -0.041  \\
\end{tabular}
\end{ruledtabular}
\label{tab:hse1}
\end{table*}

We first recall the basic properties of LaMnO$_3$.
At zero pressure and low temperature LaMnO$_3$ is a type-A AFM insulator (alternating planes of similar spins along the $c$ direction) characterized by staggered JT and GdFeO$_3$-type
(GFO) distortions, manifested by long (l) and short (s) Mn-O in-plane distances and medium (m) Mn-O
vertical ones (JT), and by the tilting of the the Mn$^{3+}$O$_6$  octahedra (see Fig. \ref{fig:1}(d)).
These structural instabilities removes the $e_g$ orbital degeneracy and stabilize an orthorhombic high-spin
($t_{2g}$)$^3$($e_g$)$^1$ orbitally ordered configuration\cite{murakami98}.
The application of hydrostatic pressure progressively quenches the cooperative JT distortions and leads
to an IMT at P$_c$=32 GPa\cite{Loa01}. The persistence of the structural distortions up to P$_c$ indicates
that the IMT is not a Mott-Hubbard type. This conclusion was initially proposed by LDA+U and Dynamical mean 
field theory studies\cite{Trimarchi05,Yamasaki06} and only very recently was confirmed by high pressure Raman
measurements\cite{Baldini11}. Baldini and coworkers\cite{Baldini11} have also reported the 
coexistence of domains of distorted and regular octahedra in the pressure range 3-34 GPa, and connected the 
onset of metallicity with the increase of undistorted MnO$_6$ octahedra beyond a critical threshold.
The concomitant presence of two distinct phases in this pressure range was confirmed by the 
X-ray absorption spectroscopy experiments of Ramos {\em et al.}\cite{Ramos11}.

Our computational study, beside providing a detailed microscopic understanding of the IMT and of
the associated competition between distorted and undistorted phases,
predicts that the onset of metallicity is associated with a FM spin transition and that the
FM-metal state develops towards a transport half-metal regime at elevated pressure ($\approx$ 100 GPa),
where the current is nearly fully spin polarized.
Our results are reported and discussed in Sec.\ref{sec3}. Before that we describe our computational 
setup, which is given  in Sec.\ref{sec2}. Finally, in Sec.\ref{sec4} we draw a summary.

\section{Methodology and Computational aspects}
\label{sec2}

All calculations were performed using the the Vienna Ab initio
Simulation Package\cite{Kresse1996, Kresse1999} (VASP) within the
Heyd, Scuseria, and Ernzerhof (HSE) hybrid density functional
scheme\cite{Heyd2003}, in which the exchange-correlation functional is
expressed as a suitable admixture of DFT and Hartee-Fock (HF):
\begin{equation}
E_{XC}^{HSE} = {\alpha}E_X^{HF,sr,\mu} + (1-\alpha)E_X^{PBE,sr,\mu} + E_X^{PBE,lr,\mu} + E_C^{PBE}
\end{equation}
where $\mu$=0.20\AA${-1}$, controls the range separation between the short-range (sr) and long-range (lr)
part of the Coulomb kernel, and $\alpha$ determines the fraction of exact HF exchange incorporated.
The parameter $\alpha$, which we set equal to 0.15, is chosen so as to provide accurate values for
the band gap, structural distortions and magnetic energies, as discussed below.

We have used a computational unit cell containing four LaMnO$_3$ formula units (i.e. 20 atoms) to 
simulate both the $P_{nma}$  and simple-cubic phase within the ferromagnetic (FM) and
type-A AFM orderings. Convergence tests on the
energy cut-off has shown that at low and intermediate pressure range
the energy difference $\Delta$E between the FM and AFM phases (our most critical quantity) changes
by less than 3 meV per formula unit (f.u.) when the energy cut-off
is increased from 300 meV to 400 meV. We have therefore chosen the value
300 meV, and made use of a 4$\times$4$\times$4 Monkhorst-Pack k-point grid
for both the exchange-correlation DFT kernel and the portion ($\alpha$=0.15)
of exact HF exchange.

We have used the following structural optimization procedure.
For each volume we have fully relaxed the
cell shape (i.e., lattice parameters $a$, $b$, and $c$ as well as
the corresponding angles between them) and all internal structural
degrees of freedom (all atomic positions). This complete
geometrical optimization allowed us to tread the structural path from the
$P_{nma}$ to the cubic phases. 
We have tested about 30 different volumes.                                    
The equation of states (Pressure-Volume curves)
were calculated in two different ways: (i) by computing the pressure
directly from the stress tensor (as automatically done by VASP) and
(ii) by applying the Birch-Murnaghan equation\cite{Birch}. Both routes
lead to the same result.

\begin{table}[h]
\caption{The band gap ($\Delta$, eV), magnetic moment ($m$,
$\mu_{B}$/Mn), and relative energy with respect to the FM ordering
calculated by HSE for different values of the mixing parameter
$\alpha$. } \vspace{0.3cm}
\begin{ruledtabular}
\begin{tabular}{lcccccc}
          & Expt.                       & HSE-35 & HSE-25 & HSE-15 & HSE-10 &  PBE   \\
$\Delta$  & 1.1$^a$, 1.7$^b$            &  3.41  &  2.47  &  1.45  &  0.75  &  0.00  \\
\emph{m}  & 3.7$^c$                     &  3.78  &  3.74  &  3.67  &  3.65  &  3.52  \\
A-AFM     &                             &  -7    &  -8    &   -24  &     3  &   171  \\
C-AFM     &                             &  156   &  182   &   198  &   368  &   564  \\
G-AFM     &                             &  161   &  192   &   208  &   428  &   899  \\
\end{tabular}
\end{ruledtabular}
\label{tab:hse2}
\begin{flushleft}
$^a$Ref.\cite{Arima},
$^b$Ref.\cite{Saitoh},
$^c$Ref.\cite{elemans71}
\end{flushleft}
\end{table}

\begin{figure}
\includegraphics[clip,width=0.5\textwidth]{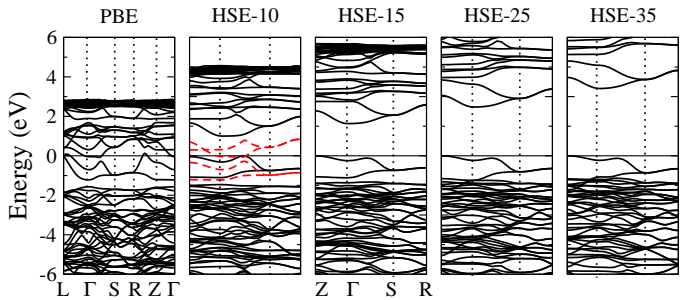}      
\caption {(Color online) PBE \& HSE band structure of LaMnO$_3$ along the Z${\Gamma}$SR (HSE) and
L${\Gamma}$SRZ${\Gamma}$ (PBE) obtained using different values of the mixing parameter at the
corresponding optimized geometry. The Fermi level is set to zero. For PBE we have included more high-symmetry
lines in order to show that both $e_g$ bands above and below $E_F$ cross the Fermi level. The PBE bands
along the Z${\Gamma}$SR are shown as (red) dashed lines in panel HSE-10.} \label{fig:band1}
\end{figure}

\subsection{Choice of the mixing factor $\alpha$: ground state properties of LaMnO$_3$}
In order to determine the HSE mixing factor $\alpha$, i.e. the fraction
of non-local HF exchange included in the hybrid exchange-correlation
functional, we have performed a set of calculations of the ground
state structural, electronic and magnetic properties of LaMnO$_3$
for different values of $\alpha$: 0 (corresponding to a purely PBE
setup), 0.10 (HSE-10), 0.15 (HSE-15), 0.25 (HSE-25) and 0.35 (HSE-35). The results, collected in
Tab.\ref{tab:hse1} (optimized geometry) and Tab.\ref{tab:hse2} (bandgap,
magnetic moment and magnetic energies), demonstrate that the best
choice is $\alpha$=0.15. For this value of the mixing parameter HSE
delivers (see Tab.\ref{tab:hse1} and Tab.\ref{tab:hse2}) (i) an insulating
bandgap, $\Delta$=1.45, well within the measured data, (ii) a
correct description of the critical cooperative Jahn-Teller (JT)
parameters Q$_2$ and Q$_3$, and (iii) a AFM magnetic ground state
in agreement with the experimental findings. Smaller mixing factors
(0 and 0.10) lead to a significant underestimation of the JT
parameters Q$_2$ and Q$_3$ (see Tab.\ref{tab:hse1}) and
to a much too small bandgap (which is actually zero in PBE, see
Tab.\ref{tab:hse2}), and to the stabilization of the FM spin
arrangement (see Tab.\ref{tab:hse2}), in contrast to the experimental
situation. This is in line with previous conventional DFT
studies\cite{Sawada97,Sawada98,Hashimoto10,Kotomin05}. We note that
by using the experimental structure a small bandgap of about 0.2 eV
is opened at PBE level, which is still too small as compared to the measured
level. Conversely, an higher $\alpha$ (0.25 and 0.35) correctly favors the
AFM ordering but overestimates the band gap (2.47 and 3.41 for $\alpha$=0.25
and $\alpha$=0.35, respectively). The value of the gap obtained for $\alpha$=0.25 
is very similar to the corresponding B3LYP value, 2.3 eV\cite{Munoz04}, and to the HSE value 
obtained using the experimental structure, 2.25 eV \cite{mlwf}.

In terms of the band dispersions the effect of the mixing parameter
is the progressively larger separation of the occupied and
unoccupied $e_g$ bands below and above the Fermi energy with
increasing $\alpha$, as shown in Fig. \ref{fig:band1}, but qualitatively
the overall bonding picture remains unchanged. In contrast, in PBE
the $e_g$ sub-bands cross the Fermi Energy and form a spurious metallic solution.                             

\begin{figure*}[t]
\includegraphics[clip,width=0.95\textwidth]{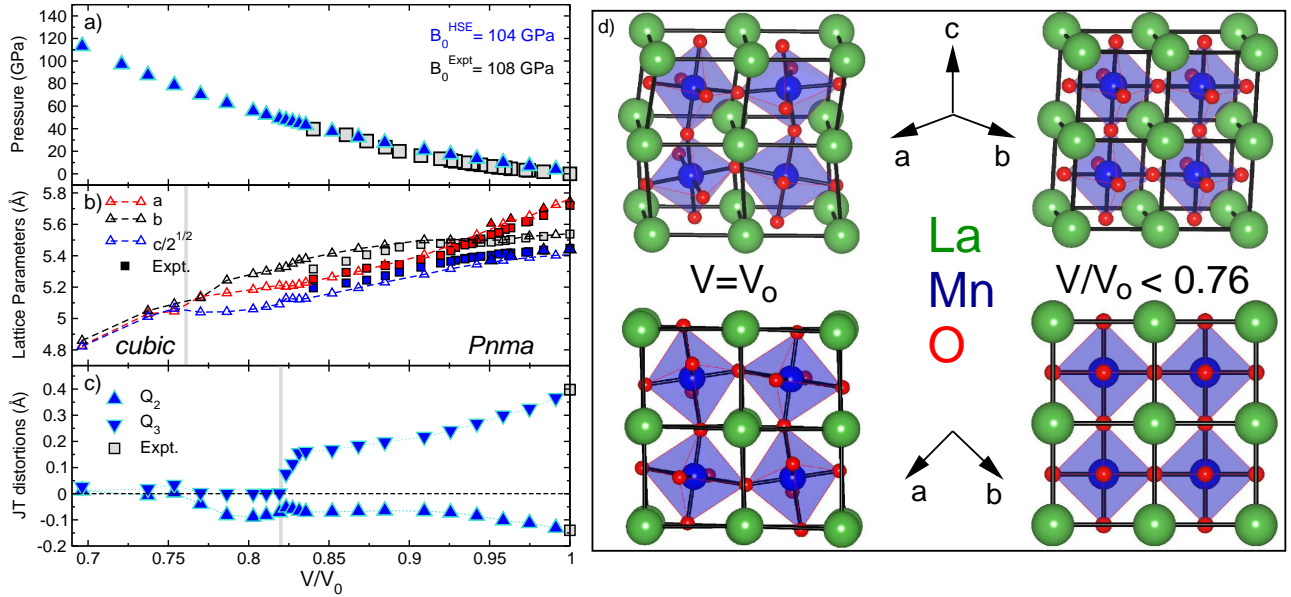}
\caption
{
(Color online) Evolution of the structural properties of LaMnO$_3$ as a function of
$\upsilon$=V/V$_0$ as predicted by HSE (triangles) and compared with the experimental data (squares)
taken from Refs. \onlinecite{Loa01, Pinsard-Gaudart01, elemans71}.
(a) Equation of state (the inset indicates the value of the bulk modulus);
(b) Structural parameters: at $\upsilon{_3}$=0.76 LaMnO$_3$ undergoes
an orthorhombic ($a\neq{b}\neq{{\sqrt{{c/2}}}}$) to cubic ($a=b={{\sqrt{{c/2}}}}$)
transition marked by the vertical line.
(c) Progressive quenching of the cooperative Jahn-Teller local modes
Q$_2=2(l-s)/\sqrt{2}$ and Q$_3=2(2m-l-s)/\sqrt{6}$ with increasing pressure,
where $l$, $s$ and $m$ indicate the long, short and medium Mn-O bondlength; the JT modes are
almost completely quenched at the onset of metallicity, marked by the vertical line at $\upsilon{_2}$=0.82
(see Fig. \ref{fig:2});
(d) Side (top) and top (bottom) view of the $Pnma$ (left, V/V$_0$$=$1) and $cubic$ (right, V/V$_0$$<$0.76)
phases of LaMnO$_3$, underlining the suppression of the JT and GFO structural distortion
in the perfect cubic phase.}
\label{fig:1}
\end{figure*}

\section{Results and Discussion}  
\label{sec3}

\begin{figure*}[t]
\includegraphics[clip,width=0.95\textwidth]{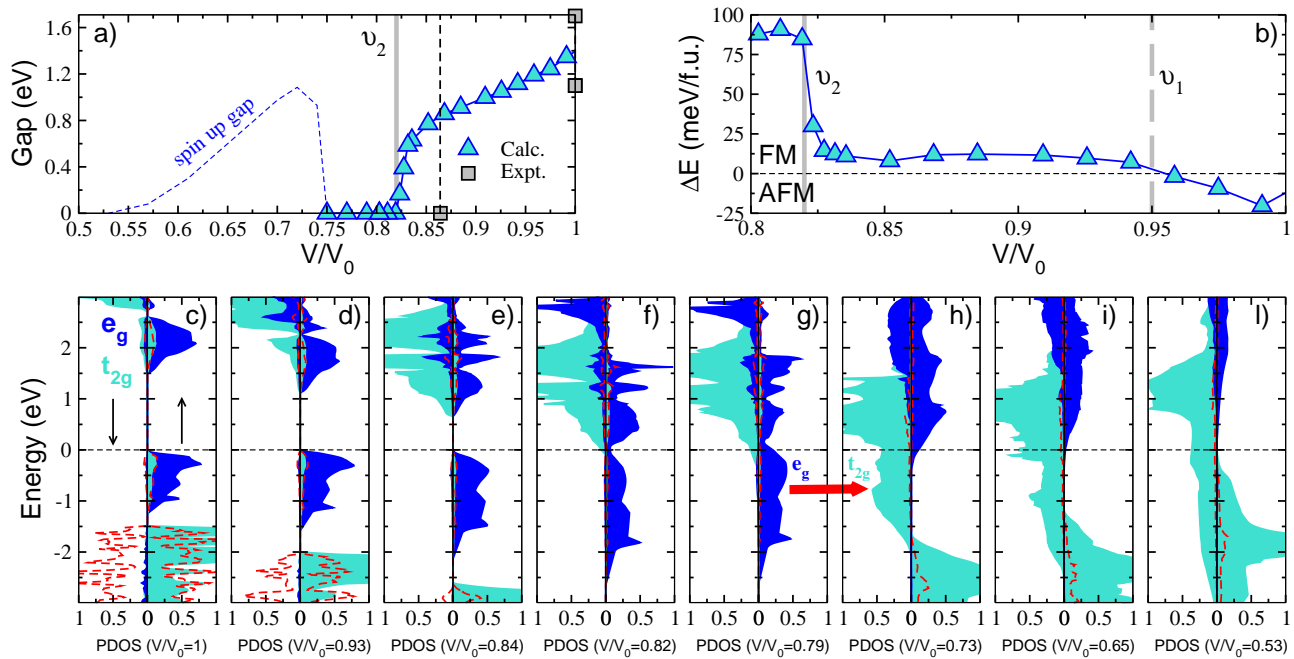}
\caption
{
(Color online) Evolution of the calculated electronic properties and magnetic ordering of LaMnO$_3$ upon pressure.
(a) Band gap: IMT at $\upsilon{_2}$=0.82 (marked by the vertical full line; the dashed line refers to the corresponding
experimental onset). Measured gaps are taken from Ref.\cite{Arima,Saitoh}.
(b) Energy difference $\Delta$E between the AFM and FM spin arrangements: AFM/FM crossover at $\upsilon{_1}$=0.95
(indicated by a vertical dashed line) and stabilization of the FM state at $\upsilon{_2}$=0.82.
(c-l) Changes in the $e_g$ and $t_{2g}$ density of states around the Fermi level with pressure.
The dashed (red) lines refer to the Oxygen $p$ states, whereas the thick (red) arrow indicates the
transfer of one electron from the $e_g$ to the $t_{2g}$ sub-bands.
}
\label{fig:2}
\end{figure*}

\begin{figure}
\includegraphics[clip,width=0.45\textwidth]{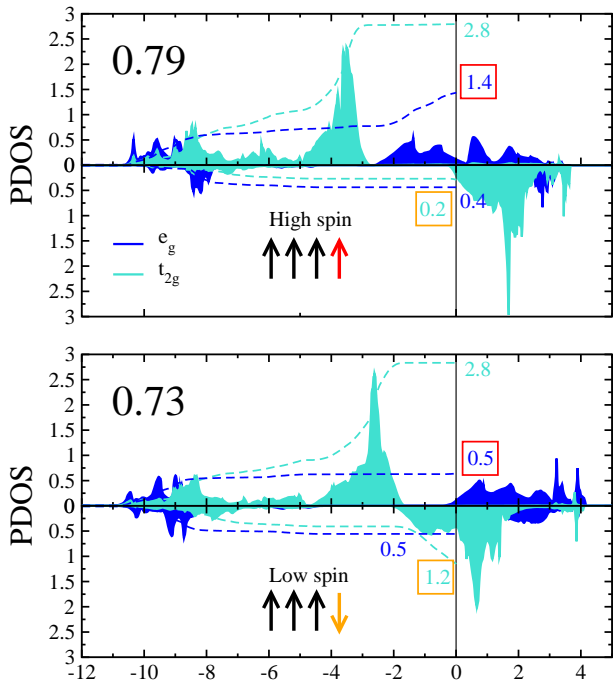}  
\caption {(Color online) Partial and integrated density of states of
LaMnO$_3$ at V/V$_0$=0.79 and V/V$_0$=0.73 representing the pressure induced
High-Spin to Low-Spin transition and associated
${e_{g}}^{\uparrow}$(V/V$_0$=0.79){$\rightarrow$}${t_{2g}}^{\downarrow}$(V/V$_0$=0.73)
electron transfer. The arrows indicate the HS and LS configuration
at V/V$_0$=0.79 and V/V$_0$=0.73, respectively, whereas the small
numbers represent the number of electrons for the specific
$lm$-channel (those in squares are those involved in the electron
transfer). } \label{fig:hl}
\end{figure}

\begin{figure}
\includegraphics[clip,width=0.45\textwidth]{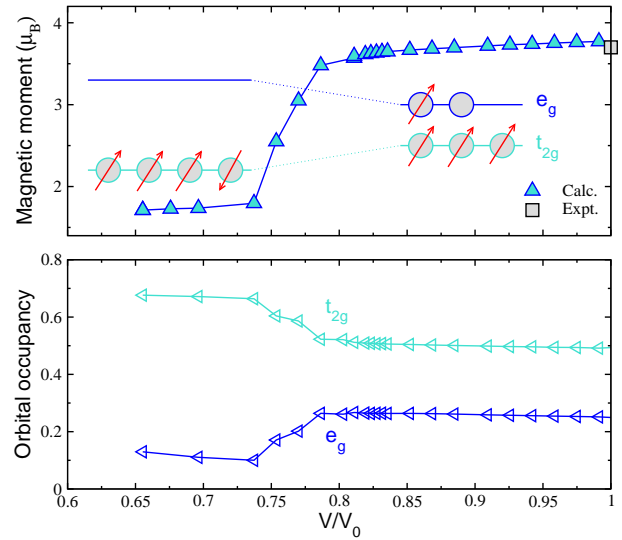}
\caption
{(Color online) 
(a) Pressure driven magnetic moment collapse and corresponding 
correlation with (b) the $e_g$ and $t_{2g}$ orbital occupancy. 
The insets in panel (a) illustrate the schematic diagrams of the HS and LS             
states at low (right) and high (left) pressure. The square indicates the experimental
magnetic moment at ambient pressure taken from Ref.\cite{elemans71}.
A similar transition has been observed in dense MnO\cite{kunes08}.
}
\label{fig:3}
\end{figure}

The progression of the structural properties of compressed LaMnO$_3$ computed by HSE 
as a function of $\upsilon$=V/V$_0$ is summarized in Fig. \ref{fig:1}, whereas
the corresponding development of the electronic and magnetic properties is shown in 
Figs. \ref{fig:2}, \ref{fig:hl} and \ref{fig:3}. 
In the pressure range 0-35 GPa, for which experimental data are available, our results are in very good 
agreement with measurements in terms of:
(i) the pressure-volume equation of states and bulk modulus B$_0$ ({B$_0$}$^{Expt}$=108 GPa, 
{B$_0$}$^{HSE}$=104 GPa, see Fig. \ref{fig:1}(a)), 
(ii) the pressure-induced changes in the structural parameters (Fig. \ref{fig:1}(b)), 
and (iii) the concurrent suppression of the JT modes Q$_2$ and Q$_3$ and the band gap at the same 
compression ($\upsilon{_2}$=0.82, slightly smaller than the experimental one, V/V$_0$=0.86, see 
Fig. \ref{fig:1}(b) and Fig. \ref{fig:2}(a)); the P=0 HSE gap opened between occupied and empty $e_g$ states, 
E$_g$=1.45 eV (Fig. \ref{fig:2}(c)), is well within the measured range, 1.1-1.7 eV\cite{Arima,Saitoh}. 
Similarly the HSE ground state values of Q$_2$ and Q$_3$ match exactly the experimental values\cite{elemans71}.

\begin{figure*}
\includegraphics[clip,width=0.95\textwidth]{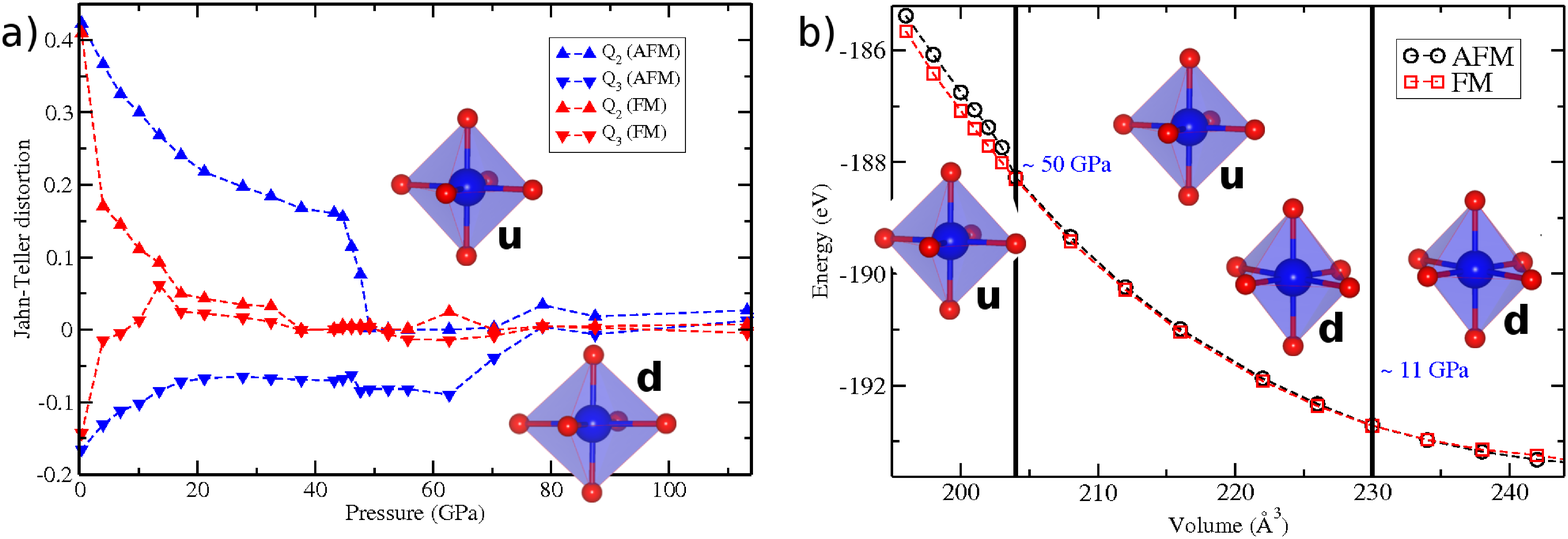}
\caption {(Color online) (a) Development of the JT quantities Q$_1$ and Q$_2$ upon pressure for the FM and AFM
ordering.
(b) Energy-Volume curve showing the coexistence of JT-distorted and undistorted phases in the
volume range $\upsilon{_2}$<V/V$_0$<$\upsilon{_1}$. This volume region is delimited by the vertical lines.
In both panels the insets represent FM-undistorted (u) and AFM-distorted (d) MnO$_6$ octahedra.
}
\label{fig:phases}
\end{figure*}

The incremental compression of LaMnO$_3$ leads to a continuous structural transformation from the P=0 
distorted $Pnma$ phase to a perfect cubic structure via a gradual quenching of the JT
modes, the rectification of the GFO tilting angles and the alignment of the $a$, $b$, and $c$ 
lattice parameters towards the same value, $\approx$ 5.1 \AA~at $\upsilon{_3}$=0.76 as outlined in 
Fig.\ref{fig:1}(b-d). The $e_g$ bands around the Fermi energy ($\rm E_F$) come progressively closer until the
gap is closed (Fig.\ref{fig:2}(c-f)). 
Concomitantly, the unoccupied $t_{2g}$ bands is pushed down in energy and ultimately crosses
the $\rm E_F$ at $\upsilon{_2}$=0.82, the onset of metallicity (see Fig. \ref{fig:2}(f)). 
At this critical volume
HSE predicts a jump in the relative stability between the AFM and FM ordering, with the 
latter becoming the most favorable one by about 90 meV/f.u., as illustrated in Fig. \ref{fig:2}(b).
At low/intermediate compressions (V/V$_0$$>$$\upsilon{_2}$=0.82) the data displayed in Fig.\ref{fig:2}(b) 
shows a strong competition between the AFM and FM phases.  HSE predicts a crossover between the AFM and FM phases
at $\upsilon{_1}$=0.95 (corresponding to a pressure of 11 GPa), below which the AFM and FM ordering 
become almost degenerate ($\Delta$E$<$12 meV/f.u.). Considering that in the FM phase the JT/GFO distortions are
almost completely inhibited (see Fig. \ref{fig:phases}), this result strongly supports the latest 
Raman\cite{Baldini11} and X-ray absorption spectroscopy\cite{Ramos11} studies reporting the formation of a mixed 
state of domains of distorted and regular MnO$_6$ octahedra in the range 13--34 GPa, which compare well with the 
corresponding theoretical pressure range, 11--50 GPa ($\upsilon{_2}$$<$V/V$_0$$<$$\upsilon{_1}$).
The coexistence of distorted and undistorted octahedra is clarified in Fig. \ref{fig:phases} where we plot 
the comparison between the evolution of the JT distortions in the FM and AFM phases (panel (a)), and the
energy-volume phase diagram (panel (b)).
In the FM phase the Q$_2$ and Q$_3$ drop down to zero at about 11 GPa (V/V$_0$=$\upsilon{_1}$=0.95), indicating that
for pressure larger than 11 GPa the MnO$_6$ octahedra are undistorted. Converesly, as already underlined,
in the AFM phase the octahedra remain distorted until V/V$_0$=$\upsilon{_2}$=0.82 (about 50 GPa). 
This different behaviour is interpreted graphically in the insets of Fig. \ref{fig:phases}(a)  which represents the 
FM-undistorted (u) and AFM-distorted (d) octahedra. Summing up, at low/high pressures LaMnO$_3$ is 
AFM-distored/FM-undistorted, whereas in the volume range $\upsilon{_2}$$<$V/V$_0$$<$$\upsilon{_1}$ these to phases
coexists in a mixed domains of distorted and undistorted octahedra (see Fig. \ref{fig:phases}(b)).

The FM transition at V/V$_0$=0.82 comes right before an high spin (HS, S=2) to low spin (LS, S=1) moment collapse, 
which is correlated with the $e_g$ and $t_{2g}$ orbital occupations as shown in Fig. \ref{fig:3}:
under compression the Mn$^{3+}$ ion retains its P=0 $(t_{2g})^{\uparrow\uparrow\uparrow}$($e_g$)$^\uparrow$ 
orbital configuration down to V/V$_0$$=$0.80, with a magnetic moment of 3.7 $\mu_B$; further compression 
yields a rapid reduction of the magnetic moment down to 1.7 $\mu_B$, due the redistribution of electrons 
within the 3$d$ shell which ultimately leads to the low-spin configuration 
$(t_{2g})^{\uparrow\uparrow\uparrow\downarrow}$($e_g$)$^0$. The HS-to-LS collapse starts to develop at
about V/V$_0$=0.8 and is fully established at exactly the same volume at which the cubic phase emerges, 
$\upsilon{_3}$=0.76. This HS-orthorhombic to LS-cubic transition is also reflected in the DOS 
(Fig. \ref{fig:2} (g-i)), whose evolution from V/V$_0$=0.79 to V/V$_0$=0.73  
clearly indicates the transfer of one electron from the $e_g$ to the $t_{2g}$
sub-bands and the subsequent realization of a nearly FM half-metallic state with a metallic minority
$t_{2g}$ band and a quasi-insulating majority channel with a residual density of $e_g$ electrons at the bottom
of the conduction band.                                                              
In order to clarify further the HS-to-LS transition we display in Fig. \ref{fig:hl} the partial and integrated
density of states in a wide energy windows associated with the $e_g$
and $t_{2g}$ bands at V/V$_0$=0.79 and V/V$_0$=0.73, which show the
transfer of one electron from the $e_g$ spin up channel at
V/V$_0$=0.79 to the $t_{2g}$ spin-down channel at V/V$_0$=0.73, which yields to the
$(t_{2g})^{\uparrow\uparrow\uparrow}$($e_g$)$^\uparrow$ to
$(t_{2g})^{\uparrow\uparrow\uparrow\downarrow}$($e_g$)$^0$ transition.

Following the classical work of Nadgorny {\em et al.}\cite{Nadgorny01}
we have analyzed the spin polarization $P_n$ associated to this high pressure nearly HM-FM state in order 
to acquire information on the spin-dependent transport properties, using the formula suggested
by Mazin\cite{Mazin99}:
\begin{equation}\label{eq:sp}
P_n = \frac{N_{\uparrow}(E_F)v_{F\uparrow}^n - N_{\downarrow}(E_F)v_{F\downarrow}^n}
{N_{\uparrow}(E_F)v_{F\uparrow}^n + N_{\downarrow}(E_F)v_{F\downarrow}^n}
\end{equation}
where $N_{\uparrow}(E_F)$, $N_{\downarrow}(E_F)$ and $v_{F\uparrow}$, $v_{F\downarrow}^n$
represent the majority and minority spin DOS and Fermi velocities, respectively, and the index $n$
refers to the different types of spin polarizations detected in spin-resolved photoemission 
measurements ($n=0$), and in ballistic ($n=1$) and diffusive ($n=2$) transport experiments.
We have computed the Fermi velocities by interfacing the VASP with the BoltzTrap code\cite{BT}, and 
obtained for V/V$_0$$=$0.70: $P_0=87\%$, $P_1=80.5\%$, and $P_2= 71\%$, and for
V/V$_0$$=$0.65: $P_0=92\%$, $P_1=87\%$ and $P_2=80\%$, values very similar to those reported for the doped 
CMR manganite $\rm La_{0.7}Sr_{0.3}MnO_3$\cite{Nadgorny01}. 
We can thus conclude, that the high pressure FM cubic phase of LaMnO$_3$ is a {\em transport} half-metal.
For denser phases (V/V$_0$$<$0.65) the majority spin band gap (from the lower laying filled $t_{2g}$ 
and the unoccupied $e_g$ band) is progressively reduced to zero at V/V$_0$$<$0.53 
(P$>$300 GPa, see Fig.\ref{fig:2}(a)).
Being the FM-tHM regime the crucial common ingredient of all CMR manganites, its realization 
in the undoped (stoichiometric) phase of the CMR parent compound LaMnO$_3$ in a wide interval of compressions, 
could help in achieving new fundamental insights into the elusive phenomena of CMR.

\section{Summary}                 
\label{sec4}

Summarizing, we have traced the development of LaMnO$_3$ upon pressure and determined a    
sequence of highly interconnected structural, electronic, and magnetic phase transitions:
(i) At ambient conditions LaMnO$_3$ posses a distorted AFM insulating state. At $\upsilon{_1}$=0.95
a competition between (distorted) AFM and (undistorted) FM configuration begin to evolve.
(ii) At the critical threshold $\upsilon{_2}$=0.82 LaMnO$_3$ undergoes a non-Mott IMT associated with
a significant reduction of the JT/GFO instabilities.
(iii) For $\upsilon{_3}$=0.76 all residual lattice distortions are suppressed and a perfect cubic
phase emerges.
(iv) At $\upsilon{_4}$=0.70 the strong crystal field splitting between $t_{2g}$ and $e_g$ drives
a magnetic moment collapse from an HS to LS tHM-FM state, manifested by a significant disproportionation
in the spin-dependent $N(E_F)$, $v_{F}$, and consequentially $P_n$. 
Our results, thus, predict that it is possible to establish a tHM-FM regime in a stoichiometric manganite
at pressure accessible by high-pressure technology.   

\section*{Acknowledgements}

Support by European Community (FP7 grant ATHENA), the Chinese Academy of Science 
(CAS Fellowship for Young International Scientists) and the National Science Foundation of 
China (NSFC Grand 51050110444) is gratefully acknowledged. 
X.-Q.C is grateful for supports from the CAS “Hundred Talents Project” 
and from NSFC of China (Grand Numbers: 51074151 and 51174188).
All calculations have been performed on the Vienna Scientific Cluster.

\end{document}